\documentclass{ws-p10x7}

\def\beq{\begin{equation}}
\def\eeq{\end{equation}}
\def\beqa{\begin{eqnarray}}
\def\eeqa{\end{eqnarray}}

\def\za{\alpha}
\def\zb{\beta}
\def\lsim{\mathrel{\raise.3ex\hbox{$<$\kern-.75em\lower1ex\hbox{$\sim$}}} }
\def\gsim{\mathrel{\raise.3ex\hbox{$>$\kern-.75em\lower1ex\hbox{$\sim$}}} }

\begin{document}
\thispagestyle{empty}

\onecolumn

\begin{flushright}
IPAS-HEP-k010\\
Aug 2000
\end{flushright}

\vspace*{.5in}

\begin{center}
{\bf  \Large Some Recent Results from the Complete Theory of SUSY without R-parity $^\star$ }\\
\vspace*{.5in}
{\bf  Otto C.W. Kong}\\[.05in]
{\it Institute of Physics, Academia Sinica, Nankang, Taipei, TAIWAN 11529}

\vspace*{.8in}
{Abstract}\\
\end{center}

We review an efficient formulation of the complete theory
of supersymmetry without R-parity, where all the admissible R-parity
violating terms incorporated. Some interesting recent results
will be discussed, including newly identified 1-loop contributions
to neutrino masses and electric dipole moments of neutron and
electron, resulted from R-parity violating $LR$ squark and slepton
mixings.

\vfill
\noindent --------------- \\
$^\star$ Talk (\# 10a-06) presented  at ICHEP 2000 (Jul 26 - Aug 2), Osaka, Japan\\
 --- submission for the proceedings.  
 
\clearpage
\addtocounter{page}{-1}

\title{Some recent results from the complete theory of SUSY without R-parity}

\author{Otto C. W. Kong}

\address{Institute of Physics, Academia Sinica, Nankang, Taipei, TAIWAN 11529\\E-mail: kongcw@phys.sinica.edu.tw}

\twocolumn[\maketitle\abstract{
We review an efficient formulation of the complete theory
of supersymmetry without R-parity, where all the admissible R-parity
violating terms incorporated. Some interesting recent results
will be discussed, including newly identified 1-loop contributions
to neutrino masses and electric dipole moments of neutron and
electron, resulted from R-parity violating $LR$ squark and slepton
mixings.
}]

\section{The Generic Supersymmetric Standard Model}
Supersymmetry without R-parity is nothing but the generic supersymmetric 
Standard Model, {\it i.e.} a theory built with the minimal superfield 
spectrum incorporating the Standard Model (SM) particles and interactions
dictated by the SM (gauge) symmetries and the idea that SUSY is
softly broken. The theory is hence generally better motivated than 
{\it ad hoc} versions of R-parity violating (RPV) theories. 
Here, however, RPV parameters come in various forms. The latter includes
the more popular trilinear ($\lambda_{ijk}$, $\lambda_{ijk}^{\prime}$,
and	$\lambda_{ijk}^{\prime\prime}$) and bilinear ($\mu_i$)
couplings in the superpotential, as well as  soft SUSY breaking
parameters of the trilinear, bilinear, and soft mass (mixing) types.
From a phenomenological point of view, there is the related notion
of (RPV) ``sneutrino VEV's". In order not to miss any plausible RPV
phenomenological features, it is important that all of the RPV
parameters be taken into consideration with {\it a priori} bias. To
emphasize the point, we call the model the {\it complete theory} of 
SUSY without R-parity.

The most general renormalizable superpotential for the supersymmetric
SM (without R-parity) can be written  as
\small\beqa
W \!\! &=& \!\varepsilon_{ab}\Big[ \mu_{\alpha}  \hat{H}_u^a \hat{L}_{\alpha}^b 
+ h_{ik}^u \hat{Q}_i^a   \hat{H}_{u}^b \hat{U}_k^{\scriptscriptstyle C}
+ \lambda_{\alpha jk}^{\!\prime}  \hat{L}_{\alpha}^a \hat{Q}_j^b
\hat{D}_k^{\scriptscriptstyle C} 
\nonumber \\
&+&
\frac{1}{2}\, \lambda_{\alpha \beta k}  \hat{L}_{\alpha}^a  
 \hat{L}_{\beta}^b \hat{E}_k^{\scriptscriptstyle C} \Big] + 
\frac{1}{2}\, \lambda_{ijk}^{\!\prime\prime}  
\hat{U}_i^{\scriptscriptstyle C} \hat{D}_j^{\scriptscriptstyle C}  
\hat{D}_k^{\scriptscriptstyle C}   ,
\eeqa\normalsize
where  $(a,b)$ are $SU(2)$ indices, $(i,j,k)$ are the usual family (flavor) 
indices, and $(\za, \zb)$ are extended flavor index going from $0$ to $3$.
In the limit where $\lambda_{ijk}, \lambda^{\!\prime}_{ijk},  
\lambda^{\!\prime\prime}_{ijk}$ and $\mu_{i}$  all vanish, 
one recovers the expression for the R-parity preserving case, 
with $\hat{L}_{0}$ identified as $\hat{H}_d$. Without R-parity imposed,
the latter is not {\it a priori} distinguishable from the $\hat{L}_{i}$'s.
Note that $\lambda$ is antisymmetric in the first two indices, as
required by  the $SU(2)$  product rules, as shown explicitly here with 
$\varepsilon_{\scriptscriptstyle 12} =-\varepsilon_{\scriptscriptstyle 21}=1$.
Similarly, $\lambda^{\!\prime\prime}$ is antisymmetric in the last two 
indices, from $SU(3)_{\scriptscriptstyle C}$. 

R-parity is exactly an {\it ad hoc} symmetry put in to make $\hat{L}_{0}$,
stand out from the other $\hat{L}_i$'s as the candidate for  $\hat{H}_d$.
It is defined in terms of baryon number, lepton number, and spin as, 
explicitly, ${\mathcal R} = (-1)^{3B+L+2S}$. The consequence is that 
the accidental symmetries of baryon number and lepton number in the SM 
are preserved, at the expense of making particles and superparticles having 
a categorically different quantum number, R-parity. The latter is actually 
not the most effective discrete symmetry to control superparticle 
mediated proton decay\cite{pd}, but is most restrictive in terms
of what is admitted in the Lagrangian, or the superpotential alone. 
On the other hand, R-parity also forbides neutrino masses in the
supersymmetric SM. The strong experimental hints for the existence of 
(Majorana) neutrino masses\cite{exp} is an indication of lepton 
number violation, hence suggestive of R-parity violation.

The soft SUSY breaking part 
of the Lagrangian is more interesting, if only for the fact that  many
of its interesting details have been overlooked in the literature.
However, we will postpone the discussion till after we address the
parametrization issue.\\[-.2in]

\section{Parametrization}

Doing phenomenological studies without specifying a choice 
of flavor bases is ambiguous. It is like doing SM quark physics with 18
complex Yukawa couplings, instead of the 10 real physical parameters.
As far as the SM itself is concerned, the extra 26 real parameters
are simply redundant, and attempts to related the full 36 parameters to
experimental data will be futile.
In SUSY without R-parity, the choice of an optimal
parametrization mainly concerns the 4 $\hat{L}_\alpha$ flavors. We use
here the single-VEV parametrization\cite{ru} (SVP), in which flavor bases 
are chosen such that : 
1/ among the $\hat{L}_\alpha$'s, only  $\hat{L}_0$, bears a VEV,
{\it i.e.} {\small $\langle \hat{L}_i \rangle \equiv 0$};
2/  {\small $h^{e}_{jk} (\equiv \lambda_{0jk}) 
=\frac{\sqrt{2}}{v_{\scriptscriptstyle 0}} \,{\rm diag}
\{m_{\scriptscriptstyle 1},
m_{\scriptscriptstyle 2},m_{\scriptscriptstyle 3}\}$};
3/ {\small $h^{d}_{jk} (\equiv \lambda^{\!\prime}_{0jk} =-\lambda_{j0k}) 
= \frac{\sqrt{2}}\,{v_{\scriptscriptstyle 0}}{\rm diag}\{m_d,m_s,m_b\}$}; 
4/ {\small $h^{u}_{ik}=\frac{v_{\scriptscriptstyle u}}{\sqrt{2}}
V_{\scriptscriptstyle \!C\!K\!M}^{\dag} \,{\rm diag}\{m_u,m_c,m_t\}$}, where 
${v_{\scriptscriptstyle 0}} \equiv  \sqrt{2}\,\langle \hat{L}_0 \rangle$
and ${v_{\scriptscriptstyle u} } \equiv \sqrt{2}\,
\langle \hat{H}_{u} \rangle$. Thus, the parametrization singles out the
$\hat{L}_0$ superfield as the one containing the Higgs. As a result,
it gives the complete RPV effects on the {tree-level mass matrices} 
of all the states (scalars and fermions) the simplest structure.
The latter is a strong technical advantage.

There are, in fact, many subtle issues involved in a consistent 
formulation of the complete theory. One has to be particularly
careful when the perspective of adding the various RPV terms to the 
MSSM is taken. Such issues will be addressed in a forthcoming
review\cite{as8}, to which the readers are referred.

\section{Fermion Sector Phenomenology}
The SVP gives quark mass matrices exactly in the SM form. For the masses
of the color-singlet fermions, all the RPV effects are paramatrized by the
$\mu_i$'s only. For example, the five charged fermions ( 3 charged leptons
+ Higgsino + gaugino ), we have
\small\beq \label{mc}
{\mathcal{M}_{\scriptscriptstyle C}} =
 \left(
{\begin{array}{ccccc}
{M_{\scriptscriptstyle 2}} &  
\frac{g_{\scriptscriptstyle 2}{v}_{\scriptscriptstyle 0}}{\sqrt 2}  
& 0 & 0 & 0 \\
 \frac{g_{\scriptscriptstyle 2}{v}_{\scriptscriptstyle u}}{\sqrt 2} & 
 {{ \mu}_{\scriptscriptstyle 0}} & {{ \mu}_{\scriptscriptstyle 1}} &
{{ \mu}_{\scriptscriptstyle 2}}  & {{ \mu}_{\scriptscriptstyle 3}} \\
0 &  0 & {{m}_{\scriptscriptstyle 1}} & 0 & 0 \\
0 & 0 & 0 & {{m}_{\scriptscriptstyle 2}} & 0 \\
0 & 0 & 0 & 0 & {{m}_{\scriptscriptstyle 3}}
\end{array}}
\right)  \; .
\eeq\normalsize
Moreover each $\mu_i$ parameter here characterizes directly the RPV effect
on the corresponding charged lepton  ($\ell_i = e$, $\mu$, and $\tau$).
This, and the corresponding neutrino-neutralino masses and mixings,
has been exploited to implement a detailed study of the tree-level
RPV phenomenology from the gauge interactions, with interesting 
results\cite{ru}.

Neutrino masses and oscillations is no doubt a central aspect of any
RPV model. In our opinion, it is particularly important to study
the various RPV contributions in a framework that takes no assumption
on the other parameters. Our formulation provides such a framework.
Interested readers are referred to Refs.\cite{ok,as1,as5,as9,AL}.

\section{Interesting Phenomenology from SUSY Breaking Terms}

The soft SUSY breaking part of the Lagrangian can be written as 
\footnotesize\beqa
&& \!\!\!\!\!\!\!\! V_{\rm soft}
= \tilde{Q}^\dagger \tilde{m}_{\!\scriptscriptstyle {Q}}^2 \,\tilde{Q} 
+\tilde{U}^{\dagger} 
\tilde{m}_{\!\scriptscriptstyle {U}}^2 \, \tilde{U} 
+\tilde{D}^{\dagger} \tilde{m}_{\!\scriptscriptstyle {D}}^2 
\, \tilde{D} +
 \tilde{L}^\dagger \tilde{m}_{\!\scriptscriptstyle {L}}^2  \tilde{L}  
 \nonumber \\
&+&
 \tilde{E}^{\dagger} \tilde{m}_{\!\scriptscriptstyle {E}}^2 
\, \tilde{E}
+ \tilde{m}_{\!\scriptscriptstyle H_{\!\scriptscriptstyle u}}^2 \,
|H_{u}|^2 
+  \Big[ \,
\frac{M_{\!\scriptscriptstyle 1}}{2} \tilde{B}\tilde{B}
   + \frac{M_{\!\scriptscriptstyle 2}}{2} \tilde{W}\tilde{W}
 \nonumber \\
&+&
\frac{M_{\!\scriptscriptstyle 3}}{2} \tilde{g}\tilde{g}
+\epsilon_{\!\scriptscriptstyle ab} \Big( \,
  B_{\za} \,  H_{u}^a \tilde{L}_\za^b 
+ A^{\!\scriptscriptstyle U}_{ij} \, 
\tilde{Q}^a_i H_{u}^b \tilde{U}^{\scriptscriptstyle C}_j 
 \nonumber \\
&+&
 A^{\!\scriptscriptstyle D}_{ij} 
H_{d}^a \tilde{Q}^b_i \tilde{D}^{\scriptscriptstyle C}_j  
+ A^{\!\scriptscriptstyle E}_{ij} 
H_{d}^a \tilde{L}^b_i \tilde{E}^{\scriptscriptstyle C}_j 
+ A^{\!\scriptscriptstyle \lambda^\prime}_{ijk} 
\tilde{L}_i^a \tilde{Q}^b_j \tilde{D}^{\scriptscriptstyle C}_k
\nonumber \\
&+& 
\frac{1}{2}\, A^{\!\scriptscriptstyle \lambda}_{ijk} 
\tilde{L}_i^a \tilde{L}^b_j \tilde{E}^{\scriptscriptstyle C}_k  \Big)
+ \frac{1}{2}\, A^{\!\scriptscriptstyle \lambda^{\prime\prime}}_{ijk}
 \tilde{U}^{\scriptscriptstyle C}_i  \tilde{D}^{\scriptscriptstyle C}_j  
\tilde{D}^{\scriptscriptstyle C}_k  
  +  \mbox{\normalsize h.c.} \Big]
\label{soft}
\eeqa\normalsize
where we have separated the R-parity conserving $A$-terms from the 
RPV ones (recall $\hat{H}_{d} \equiv \hat{L}_0$). Note that 
$\tilde{L}^\dagger \tilde{m}_{\!\scriptscriptstyle \tilde{L}}^2  \tilde{L}$,
unlike the other soft mass terms, is given by a 
$4\times 4$ matrix. Explicitly, 
$\tilde{m}_{\!\scriptscriptstyle {L}_{00}}^2$ corresponds to 
$\tilde{m}_{\!\scriptscriptstyle H_{\!\scriptscriptstyle d}}^2$ 
of the MSSM case while 
$\tilde{m}_{\!\scriptscriptstyle {L}_{0k}}^2$'s give RPV mass mixings.

Obtaining the squark and slepton masses is straight forward. The only
RPV contribution to the squark masses\cite{as5,as6} is given by a
$- (\, \mu_i^*\lambda^{\!\prime}_{ijk}\, ) \; 
\frac{v_{\scriptscriptstyle u}}{\sqrt{2}}$ term in the $LR$ mixing part.
Note that the term contains flavor-changing ($j\ne k$) parts which,
unlike the $A$-terms ones, cannot be suppressed through a flavor-blind
SUSY breaking spectrum. Hence, it has very interesting implications
to quark electric dipole moments (EDM's) and related processses
such as $b\to s\, \gamma$\cite{as4,as6,kk}. For instance, it contributes
to neutron EDM at 1-loop order, through a simple gluino diagram of the
$d$ squark. If one naively imposes the constraint for this RPV
contribution itself not to exceed the experimental bound on neutron
EDM, one gets roughly
$\mbox{Im}(\mu_i^*\lambda^{\!\prime}_{i\scriptscriptstyle 1\!1}) 
\lsim 10^{-6}\,\mbox{GeV}$, a constraint that is interesting even
in comparison to the bounds on the corresponding parameters obtainable
from asking no neutrino masses to exceed the super-Kamiokande 
atmospheric oscillation scale\cite{as4}.

Things in the slepton sector are more complicated.
The $1+4+3$ charged scalar masses are given in terms of blocks
\footnotesize\beqa
&& \widetilde{\cal M}_{\!\scriptscriptstyle H\!u}^2 =
\tilde{m}_{\!\scriptscriptstyle H_{\!\scriptscriptstyle u}}^2
+ \mu_{\!\scriptscriptstyle \za}^* \mu_{\scriptscriptstyle \za}
+ M_{\!\scriptscriptstyle Z}^2\, \cos\!2 \beta 
\left[ \,\frac{1}{2} - \sin\!^2\theta_{\!\scriptscriptstyle W}\right]
\nonumber \\
&+&  M_{\!\scriptscriptstyle Z}^2\,  \sin\!^2 \beta \;
[1 - \sin\!^2 \theta_{\!\scriptscriptstyle W}]
\; ,
\nonumber \\
&&\widetilde{\cal M}_{\!\scriptscriptstyle LL}^2
= \tilde{m}_{\!\scriptscriptstyle {L}}^2 +
m_{\!\scriptscriptstyle L}^\dag m_{\!\scriptscriptstyle L}
+ M_{\!\scriptscriptstyle Z}^2\, \cos\!2 \beta 
\left[ -\frac{1}{2} +  \sin\!^2 \theta_{\!\scriptscriptstyle W}\right] 
\nonumber \\
&+& \left( \begin{array}{cc}
 M_{\!\scriptscriptstyle Z}^2\,  \cos\!^2 \beta \;
[1 - \sin\!^2 \theta_{\!\scriptscriptstyle W}] 
& \quad 0_{\scriptscriptstyle 1 \times 3} \quad \\
0_{\scriptscriptstyle 3 \times 1} & 0_{\scriptscriptstyle 3 \times 3}  
\end{array} \right) 
+ (\mu_{\!\scriptscriptstyle \za}^* \mu_{\scriptscriptstyle \zb})
\; ,
\nonumber \\
&& \widetilde{\cal M}_{\!\scriptscriptstyle RR}^2 =
\tilde{m}_{\!\scriptscriptstyle {E}}^2 +
m_{\!\scriptscriptstyle E} m_{\!\scriptscriptstyle E}^\dag
+ M_{\!\scriptscriptstyle Z}^2\, \cos\!2 \beta 
\left[  - \sin\!^2 \theta_{\!\scriptscriptstyle W}\right] \; ; \qquad
\eeqa
and
\beqa 
\label{ELH}
\widetilde{\cal M}_{\!\scriptscriptstyle LH}^2
&=& (B_{\za}^*)  
+ \left( \begin{array}{c} 
{1 \over 2} \,
M_{\!\scriptscriptstyle Z}^2\,  \sin\!2 \beta \,
[1 - \sin\!^2 \theta_{\!\scriptscriptstyle W}]  \\
0_{\scriptscriptstyle 3 \times 1} 
\end{array} \right)\; ,
\qquad
\\
\label{ERH}
\widetilde{\cal M}_{\!\scriptscriptstyle RH}^2
&=&  -\,(\, \mu_i^*\lambda_{i{\scriptscriptstyle 0}k}\, ) \; 
\frac{v_{\scriptscriptstyle 0}}{\sqrt{2}} \; ,
\\ 
\label{ERL}
(\widetilde{\cal M}_{\!\scriptscriptstyle RL}^{2})^{\scriptscriptstyle T} 
&=& \left(\begin{array}{c} 
0  \\   A^{\!{\scriptscriptstyle E}} 
\end{array}\right)
 \frac{v_{\scriptscriptstyle 0}}{\sqrt{2}}
-\,(\, \mu_{\scriptscriptstyle \za}^*
\lambda_{{\scriptscriptstyle \za\zb}k}\, ) \, 
\frac{v_{\scriptscriptstyle u}}{\sqrt{2}} \; .
\eeqa \normalsize

We have to skip the neutral scalar part\cite{as5} here. The RPV 
contributions to the charged, as well as neutral, scalar masses and 
mixings give rise to new terms in quarks and electron EDM's\cite{as6,as7}, 
$b\to s \, \gamma$\cite{kk}, $\mu \to e\, \gamma$\cite{as7}, and
neutrino masses diagrams that have been largely overlooked\cite{as1,as5}.
The last includes diagrams corresponding to a SUSY version of the
popular Zee neutrino mass model\cite{zee}. Details are to be found
in the cited references.

\section*{Acknowledgments}
The author thanks M. Bisset, K. Cheung, S.K. Kang,
Y.-Y. Keum, C. Macesanu, and L.H. Orr, for collaborations on the subject, 
and colleagues at Academia Sinica for support. He is grateful to
the generous hospitality of  YITP, Kyoto Univ. and KEK, which 
helps to make his trip to Osaka possible.

\end{document}